\documentclass[12pt]{article}
\usepackage{graphicx}
\usepackage{amsmath}
\usepackage{amssymb}
\usepackage{hyperref}
\usepackage{braket}
\usepackage{xfrac}
\usepackage{color}

\usepackage[sort&compress,numbers,square]{natbib}
\usepackage[caption=false]{subfig}

\newcommand\pubnumber{NuPhys2016-Hostert}
\newcommand\pubdate{\today}

\def\durham{Institute for Particle Physics Phenomenology, Department of Physics,\\ Durham Univeristy, South Road, Durham DH1 3LE, United Kingdom}

\def\support{\footnote{Work supported by CNPq - Brazil.}}
\def\presenter{\footnote{Poster presenter.}}

\def\Title#1{\begin{center} {\Large #1 } \end{center}}
\def\Author#1{\begin{center}{ \sc #1} \end{center}}
\def\Address#1{\begin{center}{ \it #1} \end{center}}

\newcommand\pubblock{\rightline{\begin{tabular}{l} \pubnumber\\
         \pubdate  \end{tabular}}}
\newenvironment{Abstract}{\begin{quotation}  }{\end{quotation}}
\newenvironment{Presented}{\begin{quotation} \begin{center} 
             PRESENTED AT\end{center}\bigskip 
      \begin{center}\begin{large}}{\end{large}\end{center} \end{quotation}}

\def\beq{\begin{equation}}
\def\eeq#1{\label{#1}\end{equation}}
\def\eeqn{\end{equation}}

\def\beqa{\begin{eqnarray}}
\def\eeqa#1{\label{#1}\end{eqnarray}}
\def\eeqan{\end{eqnarray}}

\let\bar=\overbar

\def\Dslash{\not{\hbox{\kern-4pt $D$}}}
\def\dslash{\not{\hbox{\kern-2pt $\del$}}}

\def\msb{{\bar{\ssstyle M \kern -1pt S}}}

\begin{document}
\begin{titlepage}
\pubblock

\vfill
\Title{Light Sterile Neutrinos at $\nu$STORM: \\Decoherence and CP violation}
\vfill
\Author{ Peter Ballett, Matheus Hostert\support \presenter and Silvia Pascoli}
\Address{\durham}
\vfill
\begin{Abstract}
Light sterile neutrino oscillations can be partially or completely washed out at short-baseline experiments due to the breaking of neutrino production coherence. In this work we address this issue in sterile searches at $\nu$STORM, an experimental proposal for a beam of neutrinos from the decay of stored muons. We work with 3+1 and 3+2 models, the latter introducing CP violation at short-baselines. We find that decoherence effects are only relevant for sterile masses above $\Delta m^2 \gtrsim 10$ eV$^2$, and that, even in that regime, we are able to place strong appearance bounds in such clean environments. In addition, the novel signatures of CP violation in the parameter space of interest can be identified with a significance of up to $\gtrsim 3 \sigma$.
\end{Abstract}
\vfill
\begin{Presented}
NuPhys2016, Prospects in Neutrino Physics \\
Barbican Centre, London, UK,  December 12--14, 2016
\end{Presented}
\vfill
\end{titlepage}
\def\thefootnote{\fnsymbol{footnote}}
\setcounter{footnote}{0}

\section{Introduction}

The three neutrino paradigm is challenged by a few experimental anomalies at short-baselines, prompting experimental efforts to search for light sterile neutrinos. In this work we evaluate the capabilities of $\nu$STORM \cite{NuSTORMprop2013,Adey2014a} in bounding the active-sterile mixing in models with one (3+1 model) and two (3+2 model) sterile neutrinos with $\Delta m^2$ in the range of $10^{-1} - 10^3$ eV$^2$. This mass range realises the interesting scenario where oscillations could occur at the far or already at the near detector. The large masses considered can also break the condition for the coherent production of neutrino mass eigenstates, a question we address via the wave-packet formalism of \cite{Akhmedov2012}. In addition, the 3+2 model allows for new CP violating phases to become relevant at short-baselines. The sensitivity of $\nu$STORM to these novel CP violation signals for some of the 3+2 model parameter space is then evaluated.

Particularly important to this study is the fact that $\nu$STORM offers an environment with low backgrounds and well-controlled systematics. The neutrino beam is obtained from the decay of $5$ GeV pions and $3.8$ GeV muons, where muons from pion decays are stored and circulate in a race-track like storage ring. Beam monitoring devices would then drastically lower flux normalisation uncertainties, providing ideal conditions to measure neutrino-nucleus cross sections in an energy range of interest to future long-baseline experiments \cite{Soler2015}. Moreover, it allows appearance searches, for instance, to be made in a beam with very little contamination.   

\section{Short-Baseline Oscillations with Decoherence}

The finite size of the neutrino source can break the coherence of the neutrino mass eigenstates at production, leading to suppressed oscillations \cite{Akhmedov2012}. These effects can be taken into account by summing the quantum mechanical amplitude for neutrino production coherently over the finite source. With the assumption of point-like parent particles this approach has been shown to be equivalent to the usual averaging of the probability over the decay pipeline \cite{Akhmedov2012}. Whilst in the plane-wave approximation the oscillation phase is given by $I_{k j} = \exp{\left(-i \Delta m^2_{k j} L / 2E \right)}$, it can be shown that, ignoring the detector size, the coherent summation of the amplitude modifies it to
\begin{equation}
I_{k j} = \,  \frac{1}{1 - e^{-\sfrac{\Delta_p}{\xi}}} \frac{1}{1 - i\xi} \left[ 1  - e^{-\sfrac{\Delta_p}{ \xi} } e^{i  \Delta_p}\right] e^{-i \Delta},
\end{equation}
where $\Delta = \Delta m^2_{kj} L/2E$, $\Delta_p = \Delta m^2_{kj} \ell_p/2E$ and $\xi = \Delta m^2_{k j}\ell_{\text{dec}}/2E$. The size of the production region is then dictated by the parent particle decay length $\ell_{\text{dec}}$ and the pipeline length $\ell_p$. Note how the plane-wave expression is recovered when $\ell_p$ vanishes. 

The 3+1 oscillation in the short-baseline regime (taking $\Delta m^2_{31}$ and $\Delta m^2_{21}$ to zero) is effectively a two neutrino one and the appearance formula, for instance, reads
\begin{equation} \label{eq:app3+1}
P^{3+1}_{\nu_{\alpha} \to \nu_{\beta}} = 2 |U_{\alpha 4}|^2 |U_{\beta 4}|^2 \big(1 -\Re(I_{41}) \big),
\end{equation}
where $\Re$ stand for the real part. With two light steriles, the short-baseline oscillation is effectively a three neutrino one, allowing for CP violation effects to appear in an appearance channel. The interference terms carry the effective CP phase parametrized by $\eta = \arg{(U_{\alpha 5}^* U_{\beta 5} U_{\alpha 4} U_{\beta 4}^*)}$. The 3+2 appearance probability is then
\begin{align}
P^{3+2}_{\nu_{\alpha} \to \nu_{ \beta}}  =\, &2 |U_{\alpha 4}|^2 |U_{\beta 4}|^2 (1 - \Re(I_{4 1})) + 2 |U_{\alpha 5}|^2 |U_{\beta 5}|^2 (1 - \Re(I_{5 1}))  \\
+ &2 |U_{\alpha 4} U_{\beta 4} U_{\alpha 5} U_{\beta 5}|  \Re\left(e^{i \eta}\right(1 - I_{41}^* - I_{51} + I_{54}\left) \right)\nonumber.
\end{align}

\section{Experimental Sensitivities}

In this section we discuss our simulation, and then present our results for $\nu_{e} \to \nu_{\mu}$ appearance and $\nu_{\mu}$ and $\overline{\nu}_{\mu}$ disappearance. We use GLoBES \cite{Huber2007} and assume a $1.3$ kt iron-scintillator far detector (FD) at $2$ km, considered in \cite{Adey2014a}, and a $200$ tonne version of the same detector at $50$ m as a near detector (ND). These have been implemented via migration matrices \cite{Ryan}. In trying to be as comprehensive as possible in our simulation we use the full near and far detector datasets when performing our $\chi^2$ analysis, \emph{i.e.} $\chi^2_{\text{tot}} = \chi^2_{\text{ND}} + \chi^2_{\text{FD}}$. This allows us to avoid the near-to-far ratio approach and is necessary for large mass steriles, as their oscillations might impact the near detector. If that is the case, the oscillations are expected to be washed out due to the breaking of the localisation condition ($\delta x_S \ll L_{\text{osc}}$, where $\delta x_S$ is the size of the source), leading to production decoherence. The first straight section of the ring where pions are injected and muons subsequently produced is assumed to be $\ell_p = 180$ m long, which determines the maximum size of the neutrino production region.
\begin{figure}[h!]
\centering
\centering
\includegraphics[width=0.44\textwidth]{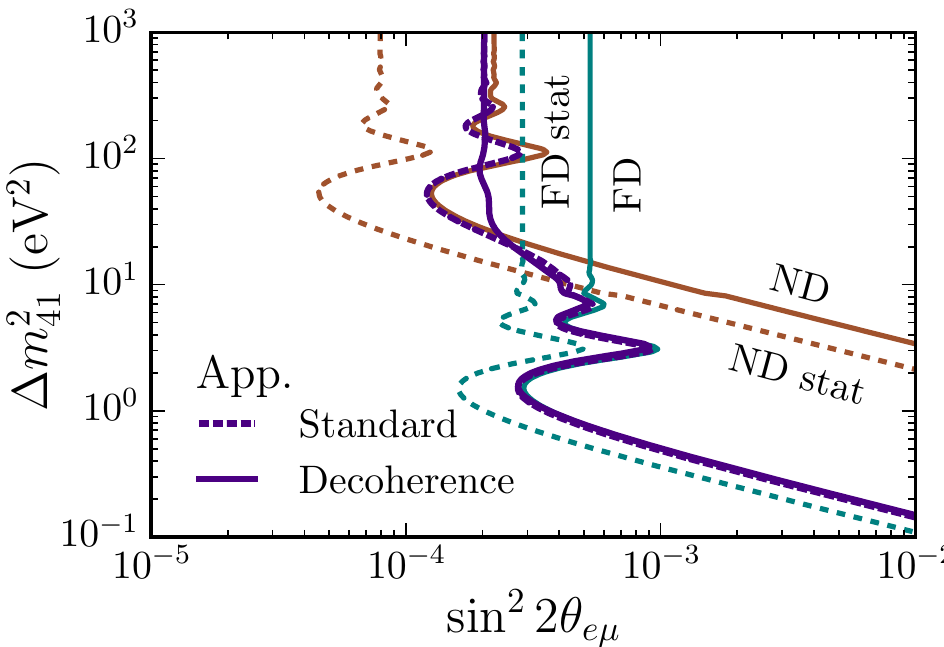}
 \quad \quad \quad
\centering
\includegraphics[width=0.44\textwidth]{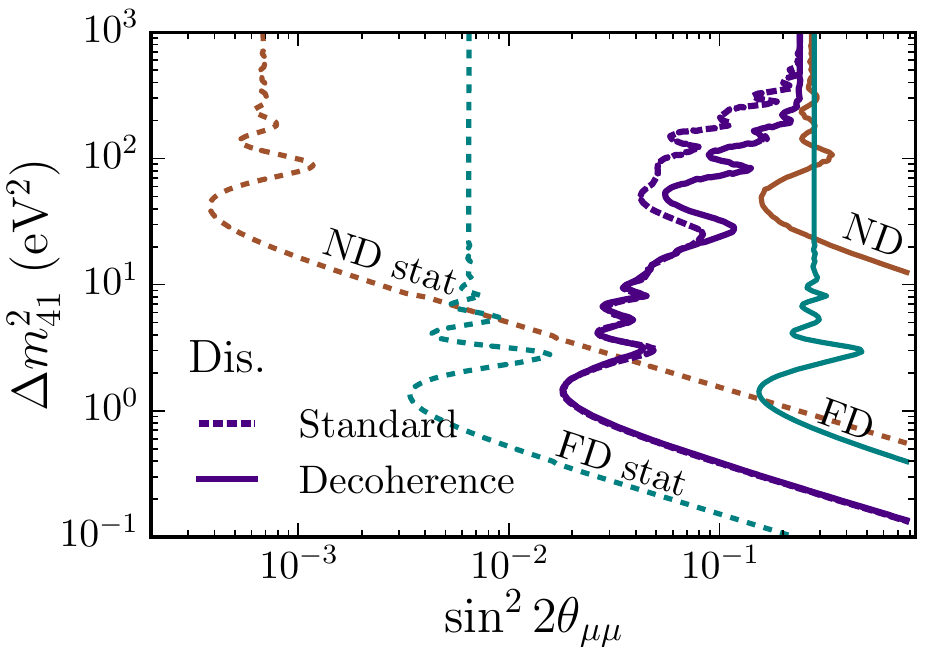}
\caption{Appearance (left) and disappearance (right) sensitivity curves at 99\% C.L. for $\nu$STORM. The brown (cyan) curve assumes only a single near (far) detector, and does not take decoherence into account. The solid (dashed) purple curve assumes the presence of both detectors with (without) decoherence.
\label{fig:3+1sens}}
\end{figure}

Finally, for our systematics we include 0.5\% flux normalisation uncertainties correlated between near and far detectors, and 0.5\% for each detector specific uncertainties. Bin dependent cross section times efficiency uncertainties are taken to be 20\%  with overall background normalisation uncertainties at the level of 35\%.

We show our results for the 3+1 model using the phenomenological parameters $\sin{2\theta_{\alpha \beta}} = 4 |U_{\alpha 4}|^2|U_{\beta 4}|^2$ in figure \ref{fig:3+1sens}. It highlights the interplay between the near and the far detectors. For low $\Delta m^2$ the near detector is not affected by oscillations and can safely measure cross sections and the flux normalisation, whilst the far detector measures the oscillation parameters. The detector roles are swapped for larger $\Delta m^2$, however, where oscillations now are washed out at the far detector, but could occur at the near detector depending on how strong the decoherence effects are. These considerations have been explored in the literature before in the context of VLENF \cite{Winter2012} and are pivotal here. Moreover, the appearance bounds are particularly interesting at large masses. In such a regime the ND, which benefits from high statistics, would be probing flavour transitions in the $\nu_{e} \to \nu_{\mu}$ channel with very low backgrounds.

\begin{figure}[h!]
\centering
\includegraphics[width=0.325\textwidth]{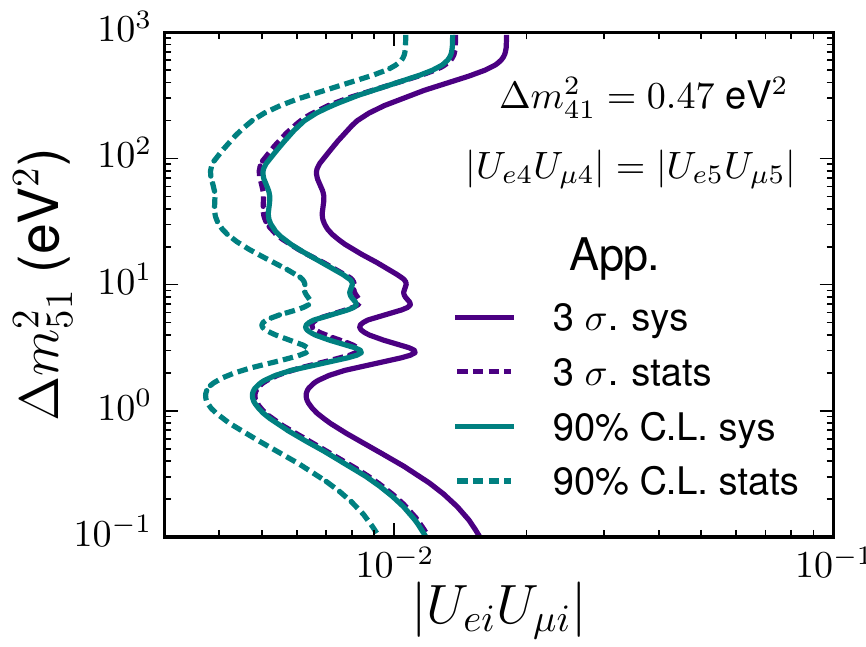}
\includegraphics[width=0.325\textwidth]{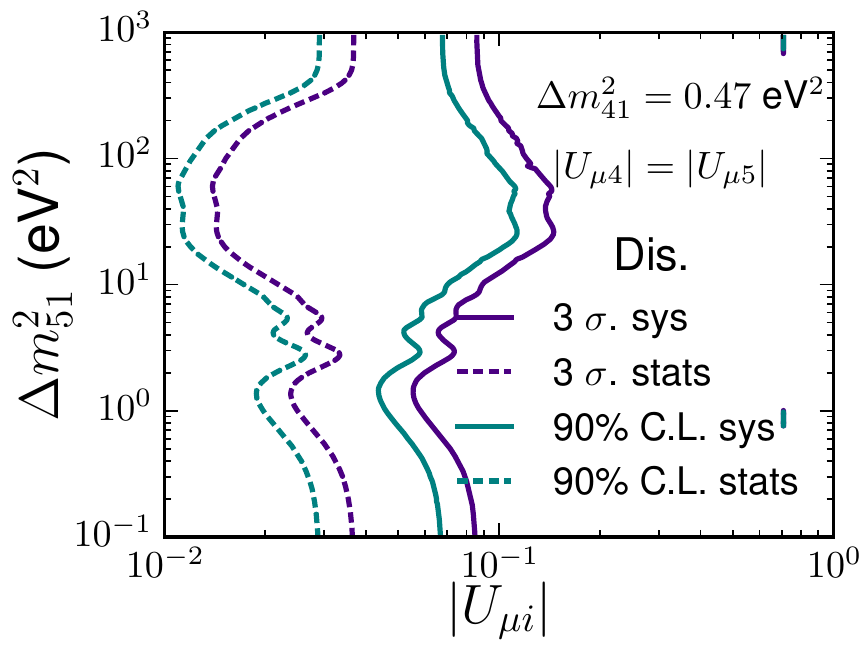}
\includegraphics[width=0.325\textwidth]{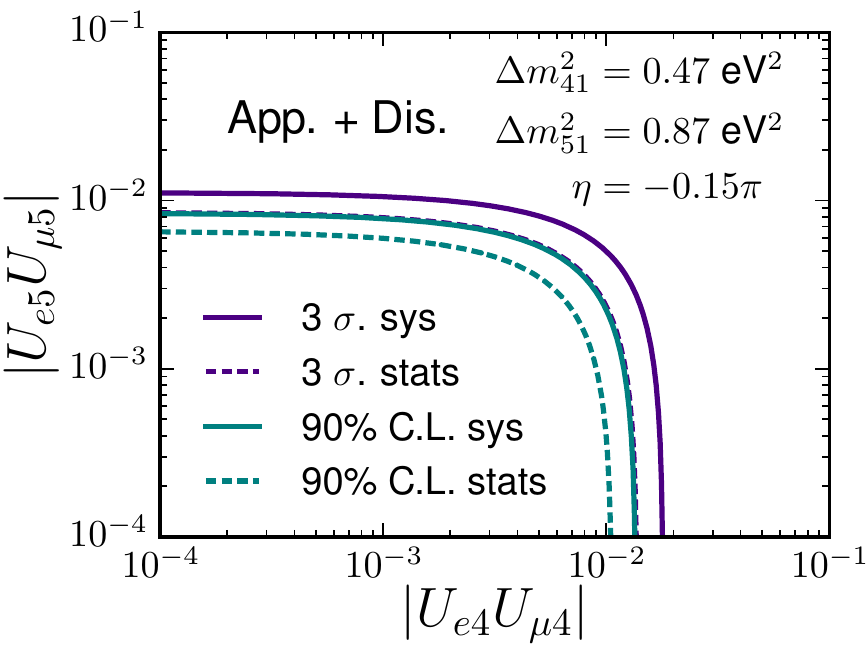}
\caption{Sensitivity of $\nu$STORM to 3+2 oscillation parameters for an appearance (left) and disappearance (center) experiment keeping $\Delta m^2_{41}$ fixed. Dashed lines show the statistical limit and solid ones include systematics. On the right we bound products of mixing matrix elements combining appearance and disappearance. We restrict ourselves to part of the parameter space as indicated in each plot. \label{fig:3+2sens} }
\end{figure}

The results for the 3+2 model are shown in figure \ref{fig:3+2sens} for specific choices of the model parameters. Note that the presence of two oscillation frequencies spoils the independence of the near and far detectors. If $\Delta m^2_{41}$ is in the $\mathcal{O}(1)$ eV$^2$ region and $\Delta m^2_{51}$ is around $\mathcal{O}(10^2)$ eV$^2$, then both near and far detectors are affected by oscillations (washed-out oscillations in the case of the ND) and the systematics cannot be safely measured at any detector. This effect is very relevant for disappearance, but not as much for appearance, which is mostly background limited.

\begin{figure}[h!] 
\includegraphics[width=0.50\textwidth]{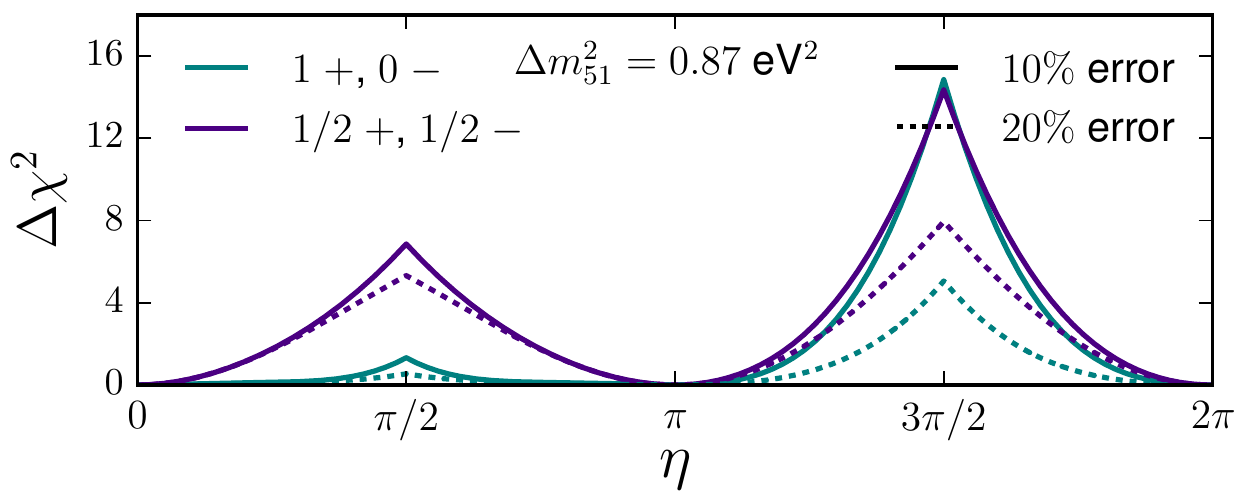} 
\includegraphics[width=0.50\textwidth]{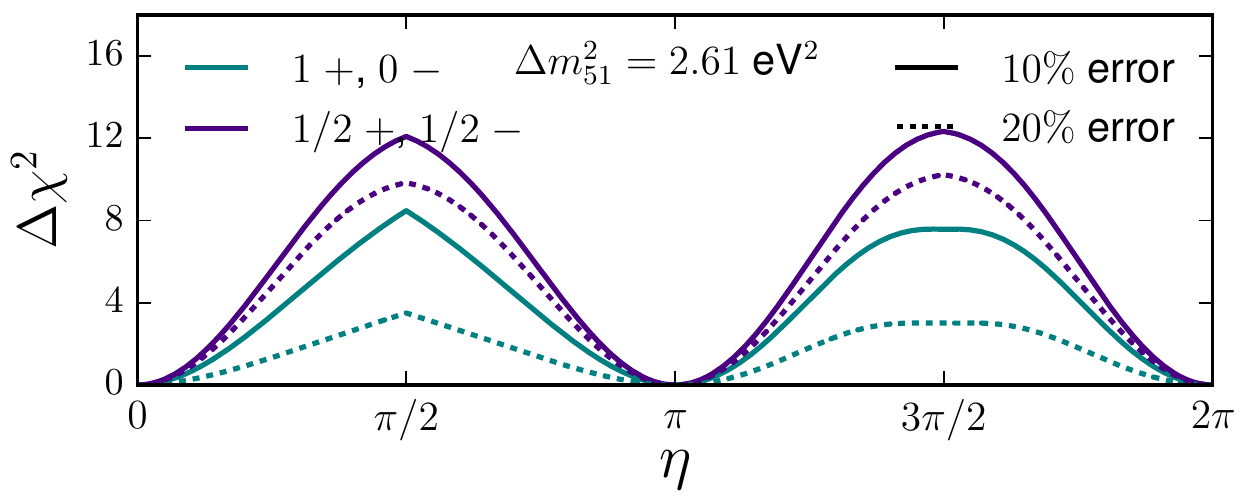} 
\caption{The sensitivity of $\nu$STORM to the CP violating phase $\eta$ when giving all other parameters errors of $10 \%$ (solid) and $20 \%$ (dashed). We take $\Delta m^2_{51} = 0.87 $ eV$^2$ on the left, using a three times as large value on the right. In cyan we show the single polarity run and in purple the run with split polarities.  \label{fig:CP1D}} 
\end{figure}

A measurement of the 3+2 model parameters assuming CP conservation is expected to have uncertainties between 10\% and 20\% for the following choice of true parameters: $\Delta m^2_{41} = 0.47 $ eV$^2$, $\Delta m^2_{51} = 0.87 $ eV$^2$, $|U_{e4}|=0.13$, $|U_{e5}|=0.14$, $|U_{\mu4}|=0.15$ and $|U_{\mu5}|=0.13$. Using this fact, we evaluate the sensitivity of $\nu$STORM to the effective CP phase $\eta$ with the same choice of true parameters, later increasing $\Delta m^2_{51}$ by three times its value. Our results are shown in figure \ref{fig:CP1D} using a single polarity run (collecting $\pi^+$'s and storing $\mu^+$'s) and splitting the runtime of $\nu$STORM in two opposite polarities.

\section{Conclusions}

The low backgrounds and low systematics at a facility like $\nu$STORM provide an ideal way to constrain the existence of light sterile neutrinos. We have shown that decoherence effects are small and that appearance bounds can be very robust, applying to sterile masses much above the few eV. A second light sterile neutrino can spoil the interplay between near and far detectors and weaken the disappearance bounds dramatically, but it reinforces the strength of appearance searches. Finally, the effective CP violation in the 3+2 model can be seen at $\nu$STORM at more than 3$\sigma$ for some parts of the parameter space if it is maximal.

\bibliography{Custom.bib}

\end{document}